\documentclass{icrc2009}

\usepackage{graphicx}   
\usepackage{caption}    
\usepackage[font=footnotesize]{subfig} 
\usepackage{fixltx2e}
\usepackage{hyperref}  
\usepackage{url}
\newcommand{\shorttitle}[1]%
{\markboth{Proceedings of the 31\MakeLowercase{$^{st}$} ICRC, {\L}\'{o}d\'{z} 2009}{#1} }
\newcommand{\etal}{\MakeLowercase{\textit{et al. }}} 

\begin{document}
\title{Measurement of the Cosmic Ray B/C Ratio\\ with the AMS-01 Experiment}
\author{\IEEEauthorblockN{Nicola Tomassetti\IEEEauthorrefmark{1}, on behalf of the AMS-01 Collaboration}
                            \\
\IEEEauthorblockA{\IEEEauthorrefmark{1}
INFN-Sezione di Perugia and Universit\`a degli Studi di Perugia, 06100 Perugia, Italy
}

}
\shorttitle{Nicola Tomassetti \etal -- Cosmic Ray B/C Ratio With AMS-01}
\maketitle

\begin{abstract}
The Alpha Magnetic Spectrometer (AMS) is a particle physics detector designed for a high precision measurement of cosmic rays in space.
AMS phase 2 (AMS-02) is scheduled to be installed on the ISS for at least three years from September 2010.
The AMS-01 precursor experiment operated successfully during a 10-day NASA shuttle flight in June 1998. The orbital inclination was 51.7$^\circ$ at a geodetic altitude between 320 to 380 km.
Nearly 200,000 Z$>$2 nuclei were observed by AMS-01 in the rigidity range 1-40 GV. Using these data, it is possible to investigate the relative abundances and the energy spectra of the primary cosmic rays, providing relations with their sources and propagation processes.
Preliminary results on the B/C ratio in 0.4-19 GeV/nucleon kinetic energy will be presented.

\end{abstract}


\section{Introduction}

The boron to carbon nuclear ratio is very important for understanding the
cosmic ray propagation in the Galaxy.
Boron and carbon are respectively a pure secondary and a pure primary cosmic ray species.
Since B is directly produced by C spallation in the interstellar medium, their relative abundance is
sensitive to the traversed matter from the source, and the B/C energy distribution is an
interesting quantity to study cosmic rays diffusion and interactions in the galactic medium.
A precise measurement of the B/C ratio constraints the propagation parameters and, as a
consequence, provides a better understanding of the astrophysical processes at the sources.
The AMS-01 measurement of the B/C ratio is presented here in the kinetic energy range 0.4-19 GeV/n.

\section{The AMS-01 detector}

The AMS-01 spectrometer was composed of a cylindrical permanent magnet, a silicon microstrip tracker,
time of flight (TOF) scintillator planes, a Cerenkov counter and anti-coincidence counters \cite{ref_AMS01}.
The magnet (inner diameter $1.1 m$) provided a central dipole
field with an analyzing power $BL^2=0.14 Tm^2$. Six layers of double-sided silicon microstrip tracker
measured the trajectory of charged particles with an accuracy
of $10 \mu m$ ($30 \mu m$) in the (non)bending coordinates, as well as providing multiple energy loss measurements.
The TOF system had two orthogonal planes at each end of the magnet, and measured the particle
transit time with an accuracy of $\sim$120 psec; they also yielded energy loss measurement up to $|Z|$=2.
A layer of anti-coincidence scintillation counters lined the inner surface of the magnet.
A thin carbon layer were used as a shield to absorb low energy particles.
\begin{figure}[!htbp]
\centering
  \includegraphics[width=0.45\textwidth]{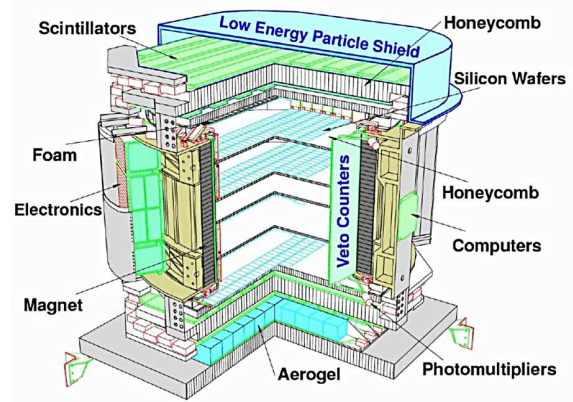}
\caption{The AMS-01 detector}\label{ams01}
\end{figure}

The detector response was simulated under a \texttt{GEANT3}-based software
\cite{ref_GEANT}. The effects of energy loss, multiple scattering, nuclear interactions and decays
were included in the code, as well as efficiencies, resolutions and reconstruction algorithms.

\section{Event reconstruction and selection}
The main physical characteristics of a cosmic ray particle traversing the detector
are the arrival direction ($\theta$,$\phi$), the particle identity and its
kinetic energy or rigidity. These quantities were reconstructed combining the independent
measurements provided by the various sub-detectors.
The particle rigidity $R=pc/Ze$ was given by the deflection of the particle
trajectory reconstructed from hits in at least four out of the six tracker planes.
The velocity $\beta=v/c$ was determined from the transit time in
the TOF along the track lenght.
The particle charge $|Z|$  was obtained by the analysis of the multiple measurements of energy loss,
as described in detail in the next section.
\\

In order to reject poorly reconstructed particles, a set of quality cuts was applied to the sample:
\begin{enumerate}

\item particles were required to be measured by fully efficient TOF scintillator paddles; this rejects events
passing through a single TOF paddle providing unreliable timing information
\item tracks fitted with high ${\chi}^{2}$ were removed according to a rigidity dependent cut-off value on the ${\chi}^{2}$;
\item tracks with poor \emph{half rigidity agreement}\footnote{For a given track, the half rigidities $R_1$ and $R_2$ are two different measurements of the particle rigidity; $R_1$ ($R_2)$ was obtained with hits from the upper (lower) silicon planes.} were also removed, according to rigidity-dependent criteria;
\item consistency between independent $\beta$ and R measurements was required,
being $\beta$ (particle velocity) and R (rigidity) obtained by two distinct subdetectors;
\item the absence of hits was required close to the track extrapolation on missing layers,
namely on tracker layers with no hit belonging to the reconstructed track.

\end{enumerate}
For the present work we considered data collected during these two periods:
(a) \emph{MIR-docking phase}, 4 days while the shuttle was docked with the MIR space station, and
(b) \emph{post-docking phase}, 2 days pointing at fixed directions ($0^0$, $20^0$ and $45^0$ with respect to the zenith).

Data collected while AMS-01 was passing in the region of the South Atlantic Anomaly were excluded
(latitude: $-47^0 \div -4^0$, longitude: $-85^0 \div -5^0$)
and only downward-going particles (in the AMS-01 reference frame)
were considered, within a restricted acceptance of $32^\circ$.
Approximately 75,000 nuclei from Lithium to Oxygen were selected with the
above mentioned criteria.

\section{Particle identification}

\begin{figure}[!h]
\centering
  \includegraphics[width=0.45\textwidth]{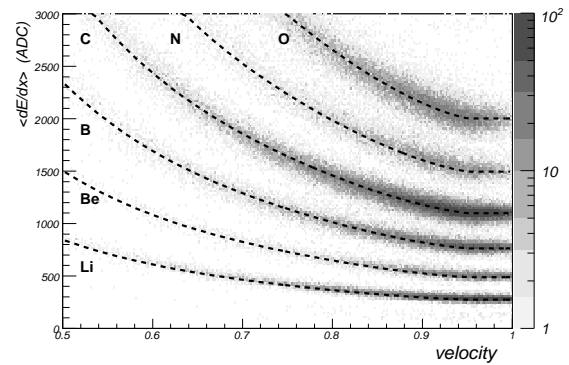}
\caption{
Signal amplitude (ADC) of the mean energy loss in the silicon tracker vs velocity.
Nuclear families fall into distinct charge bands.
MPVs of the $P_{Z}^{k}(x_{k},\beta)$ functions are superimposed for Z=3 to 8 (dashed lines).
}\label{dedxvsbeta}
\end{figure}

A nuclear species is defined by its charge Z. In this analysis
the charge of each species is determined by the energy losses recorded in the
silicon layers, due to the limited dynamical response of the
TOF scintillators.
The ionization energy generated by a charged particle in a silicon microstrip detector is
collected by a \emph{cluster} of adjacent strips. Tracker clusters were recognized online
and then re-processed with the reconstruction software.
A multi-step procedure of normalization of the cluster amplitudes was performed.
The method accounts for saturation effects, electronics response, particle inclination and velocity dependence
of the energy loss.
The charge identification algorithm, applied to the corrected signals, was based on the \emph{maximum likelihood} method
which determines the best Z value corresponding to the maximum value of the \emph{log-likelihood function}:
\begin{equation}\label{equa1}
L(Z) = \log_{10}\{\prod_{k=1}^{k=6}P_{Z}^{k}(x_{k},\beta)\}
\end{equation}
The k-index runs over the silicon layers, $x_{k}$ are the corrected
cluster amplitudes as observed on k-th layer.
The $P_{Z}^{k}(x_{k},\beta)$ functions have to be viewed as the probabilities of a given charge
Z with velocity $\beta$ to produce a signal $x_{k}$ on the k-th layers. These probability density
functions were estimated from a clean reference sample of flight data. For a fixed $\beta$,
they have gaussian-like shape; their dependence on $\beta$ is shown on Fig.\ref{dedxvsbeta}.
Due to inefficiencies in charge collection, 
some energy losses may produce charge responses which do not carry reliable
information on the particle charge. For this reason, not all
the six layer clusters were used to determine Z. 
Firstly, a selection of the tracker clusters was done according to (1) reasonable strip occupancy levels
(clusters containing noisy or dead channels were removed) and (2) cluster morphology (double-peaked clusters
and single-strip clusters were removed).
Finally the \emph{best} set of three clusters given by
$\Omega=\{k_1,k_2,k_3\}$ was recognized. $\Omega$ was provided by the maximum likelihood method as a
parameter as well as Z.
Hence, the actual \emph{log-likelihood} function used was:
\begin{equation}\label{equa2}
L(Z,\Omega) = \log_{10}\{P_{Z}^{k_1}\cdot P_{Z}^{k_2}\cdot P_{Z}^{k_3} \}
\end{equation}
where Z ranges from 3 (lithium) to 8 (oxygen) and $\Omega$ runs over all the
$\{k_1,k_2,k_3\}$ 3-fold combinations of the tracker signals.
It should be noted that, although the final Z estimation was provided by three tracker clusters,
all the selected clusters were processed in Eq.\ref{equa2}.

\begin{figure}[!htbp]
\centering
  \includegraphics[width=0.45\textwidth]{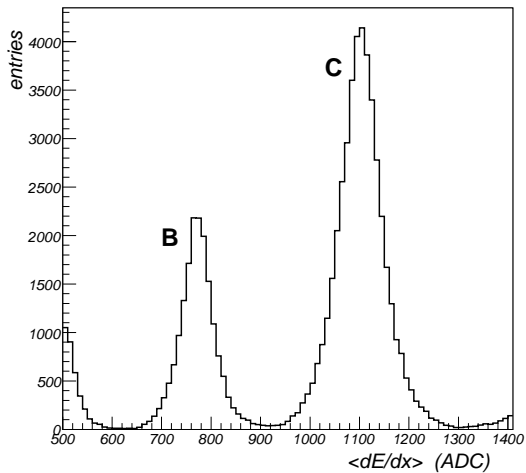}
\caption{
Charge histograms in the B and C region.
The signal amplitudes of Fig.\ref{dedxvsbeta} are here equalized to $\beta\equiv 1$.
Boron and Carbon families fall in distinct charge peaks.
}\label{dedx_bc}
\end{figure}

\section{Charge contamination}

To obtain the differential fluxes impinging the detector, the influence of charge contamination
must be accounted for.
Each nucleus of charge $3\leq Z\leq 8$ produces a charge estimation $\hat{Z}$ which is
related to its \emph{true} impinging charge $Z$ by using a set of coefficients $F_{Z}^{\hat{Z}}$.

The 6$\times$6 matrix $\|F\|$ is diagonally-dominated, and each off-diagonal element
$F_{Z}^{\hat{Z}}$ represents the probability of a nuclear species $Z$ to be
mis-identified as $\hat{Z}$ due to interactions in the detector material and fluctuations
of the energy loss:
\begin{equation}\label{equa3}
F_{Z}^{\hat{Z}}=P(\hat{Z}|Z)
\end{equation}

Two different contributions produce a charge migration $Z\rightarrow \hat{Z}$ :
\\

\begin{enumerate}
  \item after interacting in TOF material, an incoming nucleus Z may fragment and physically turn into $\hat{Z}$.
These events were typically removed by the anticoincidence veto. However, a fraction of them produces
a clean track on the tracker, passing trigger and selection. Since $\hat{Z}$$<$$Z$, the corresponding matrix is triangular.
  \item due to energy loss fluctuations and broadening of the charge response, the overlapping
of the measured $dE/dx$ distributions produces an additional $Z\rightarrow \hat{Z}$ migration probability.
 This effect is tipically symmetric and in most cases $\hat{Z}=Z\pm 1$.
\\

\end{enumerate}

Both the effects were studied with a Monte Carlo simulation where nuclear interactions and
detector responses were reproduced. An upper limit to the contribution (2) was also estimated by the data.
In principle, by the exact knowledge of $F_{Z}^{\hat{Z}}$ coefficients, it would be possible to correct the
measured abundances $\hat{Z}$ to get the corrected ones (e.g. inverting the $\|F\|$ matrix).
Since the overall distortion is estimated to be $2\% \div 6\%$ for the ratio B/C (i.e. smaller than
the statistical uncertainties), this effect was included in the systematic errors.

\section{Flux calculation}

Detector counts are relate to the differential energy flux as a function of the kinetic
energy per nucleon $E_{kn}$ according to the following relation:

\begin{equation}\label{equa4}
\Phi^{Z}(E_{kn})=\frac{N^{Z}(E_{kn})}{A^{Z}(E_{kn})\cdot T^{Z}(E_{kn})\cdot \Delta E_{kn}}
\end{equation}
where $N^{Z}(E_{kn})$ is the number of the particles detected after selection in the energy bin around $E_{kn}$,
$A^{Z}(E_{kn})$ is the detector acceptance in the selected angular range, $T^{Z}(E_{kn})$ is the exposure time and $\Delta E_{kn}$ is
the width of the energy window around $E_{kn}$.

The acceptance $A^{Z}(E_{kn})$ is the convolution of the geometrical factor with the energy dependent
efficiency $\epsilon$, which includes
(1) trigger efficiency, (2) reconstruction efficiency and (3) selection efficiency.
A Monte Carlo estimation of the acceptance was performed according to the formula \cite{ref_Sullivan}:
\begin{equation}\label{equa5}
A^{Z}(E_{kn})=A^{0}\frac{N_{D}^{Z}(E_{kn})}{N_{G}^{Z}(E_{kn})}
\end{equation}
where $N_{G}^{Z}(E_{kn})$ is the number of Z-charged nuclei generated in the energy bin $E_{kn}$
and $N_{D}^{Z}(E_{kn})$ the number of those selected.
Trajectories were generated from an ideal plane placed above the detector,
and $A^0=47.78$ $m^2sr$ is the top plane geometrical acceptance.
To calculate $A^{B}$ and $A^{C}$, $2\cdot 10^{8}$ boron and carbon trajectories were generated
in $2\div 250$ GeV/c momentum range, according to an isotropic distribution and a
power law momentum distribution with spectral index $\gamma=-1$.

The geomagnetic cut-off, varying in the 1-15 GV interval along the AMS-01 orbit, introduces different
distortions of the measured energy spectrum for different nuclear species.
The vertical cut-off was calculated in the
eccentric dipole approximation \cite{ref_Smart}:
\begin{equation}\label{equa6}
R_{vc}=15.0\cos^{4}(\lambda)/R^{2}
\end{equation}
where $\lambda$ is the geomagnetic latitude and R is the distance from the dipole center in terrestrial radii units.
To remove any distortion due to geomagnetic field, only particles with rigidity greater
than $1.2 R_{vc}$ were accepted.
This means that, for a given species Z in the energy window $\Delta E_{kn}$, only some geomagnetic regions
were accessible by AMS-01 during the orbit.
The exposure time $T^{Z}(E_{kn})$  is then defined as the overall time spent by AMS-01 in the regions
above cut-off for a certain species Z in the energy window around $E_{kn}$.
These $T^{Z}$ were also corrected for the trigger deadtime fraction.
\\

The B/C ratio is then derived as the ratio of the differential fluxes

\begin{equation}\label{equa7}
B/C = \Phi^{B}(E_{kn})/\Phi^{C}(E_{kn})
\end{equation}
Two distinct measurements were made:
(a) in the $0.4\div 2.0$ GeV/n range, the $E_{kn}$ was directly obtained from $\beta$ measurement;
(b) between $2.0$ GeV/n and $19.0$ GeV/n the $E_{kn}$ was calculated by the rigidity from tracker.
In the latter case, since the conversion $R\rightarrow E_{kn}$ requires the particle mass,
an assumption on the B and C isotopic compositions is needed\footnote{
Rigidity is the quantity directly measured by the tracker; it was translated into kinetic energy per nucleon according to the relation: $E_{kn}(R)=\sqrt{R^{2}(Z/A)^{2}+M_{p}^{2}}-M_{p}$, being $M_p$= nucleon mass and A= mass number.}.
According to various measurements and predictions (e.g. \cite{ref_SMILI, ref_Voyager, ref_GALPROP}),
the cosmic ray boron is composed of a non-negligible fraction of $^{10}B$. We considered B as a $^{10}B+^{11}B$
isotopic mixture and assumed a $^{10}B$ isotopic fraction as
\begin{equation}\label{equa8}
Y_{B}=^{10}B/(^{10}B+^{11}B)=0.35\pm0.15
\end{equation}
whereas the carbon flux is assumed to be composed of pure $^{12}C$. The uncertainty on $Y_{B}$ is consequently
transposed as additional errors on the B/C ratio ($3\%\div5\%$ relative errors).

\section{Results and discussion}
\begin{figure}[htbp]
\centering
  \includegraphics[width=0.55\textwidth]{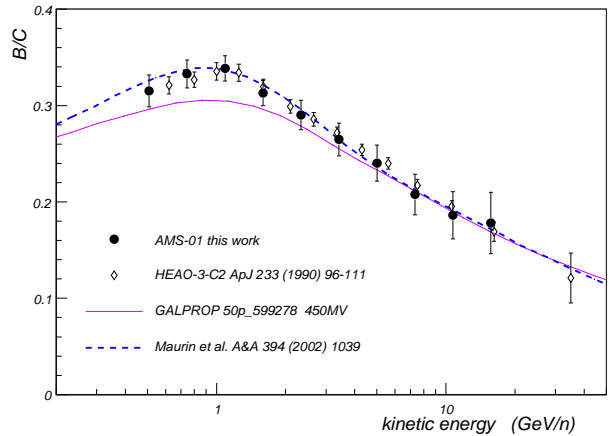}
\caption{B/C ratio from this work (filled circles), from HEAO-3-C2 experiment (open diamonds) and from two diffusion models (solid and dashed lines).}\label{bc_ratio}
\end{figure}

The results for B/C ratio are presented in Fig.\ref{bc_ratio}.
The ratio B/C is calculated in ten energy intervals between 0.4 and 19.0 GeV/n.
The energy range was decided according to restrictions due to signal saturations
(at lower energies) and requirement of reasonable statistics (at higher energies).
The binning is uniform in logarithm scale; for each data point the reference value
$\langle E_{kn} \rangle$ is defined as the geometric mean of the bin margins.
The results are compared with HEAO-3-C2 data \cite{ref_HEAO} collected
from October 1979 and June 1980.
Both the detectors operated in space during periods of similar solar activity and polarity;
altitude and inclination of the orbit were also comparable.
The two measurements are in substantial agreement.
A theoretical prediction, calculated with the \texttt{GALPROP} numerical software \cite{ref_GALPROP}
(conventional diffusion model modulated with $\Phi=450 MV$ \cite{ref_Gleeson}) is also reported.
The dashed line is referred to a diffusion model (D. Maurin \etal \cite{ref_Maurin}) where
the acceleration and diffusion processes are considered in terms of a rigidity-dependent diffusion coefficient
\begin{equation}\label{equa9}
 K(R)=K_{0}\beta \left( \frac{R}{1GV} \right) ^{\delta}
\end{equation}
and a power-law distribution $R^{-\alpha}$ is assumed at the source;
$\delta=0.6$ and $\alpha=2.2$ are used; the solar modulation is described in the force-field
approximation using $\Phi=500MV$, consistent with the period of the AMS-01 flight.

Both statistical and systematic uncertainties are represented in the Fig.4.
Most of the systematic uncertainties arising from the detector response cancel into the B-C ratio. Differences in the trigger and selection efficiencies of the two species are expected from the Z dependence of the delta-rays production and fragmentation effects in the detector material. Deep investigation on these effects has been undertaken with GEANT4 and FLUKA \cite{ref_FLUKA} based simulations with the aim of obtaining additional estimations of acceptances and their related systematics.

Finally, since a considerable amount of light nuclei were recognized in the same
energy range, one may expect to extend this analysis to other nuclear species.

\section{Aknowledgements}
The support of INFN, Italy, ETH-Zurich, the University of
Geneva, the Chinese Academy of Sciences, Academia Sinica
and National Central University, Taiwan, the RWTH Aachen,
Germany, the University of Turku, the University of Technology
of Helsinki, Finland, the US DOE and MIT, CIEMAT,
Spain, LIP, Portugal and IN2P3, France, is gratefully acknowledged.
The success of the first AMS mission is due to many individuals
and organizations outside of the collaboration. The support
of NASA was vital in the inception, development and operation
of the experiment. Support from the Max-Planck Institute for
Extraterrestrial Physics, from the space agencies of Germany
(DLR), Italy (ASI), France (CNES) and China and from CSIST,
Taiwan also played important roles in the success of AMS.
\\

\end{document}